# FOREWORD
# LEX REFORMATICA: FIVE PRINCIPLES OF POLICY REFORM FOR THE TECHNOLOGICAL AGE

*Sonia K. Katyal*[†]

## I.   INTRODUCTION

Almost twenty five years ago, our beloved former colleague Joel Reidenberg penned an article that argued that law and government regulation were not the only source of authority and rulemaking in the Information Society.[1] Rather, he argued that technology itself, particularly system design choices like network design and system configurations, can also impose similar regulatory norms on communities.[2] These rules and systems, he argued, comprised a Lex Informatica—a term that Reidenberg coined in historical reference to "Lex Mercatoria," a system of international, merchant-driven norms in the Middle Ages that emerged independent of localized sovereign control.

This work was an iconic piece of literature. For Reidenberg, there were clear parallels between merchants' travels across borders and today's users, traveling across the information superhighway, surpassing local sovereignty and confronting a tangle of conflicting regulations along the way.[3] It is no surprise that this landmark article was published in 1998, the same year that Congress passed the Digital Millennium Copyright Act (DMCA), one of the first attempts of legislators to address the onset of the digital era.[4] And Reidenberg's theory provided the backbone for yet another tour de force that defined the relationship between law and the internet, Lawrence Lessig's *Code and Other Laws of Cyberspace*, published shortly afterward.[5]

---

DOI: https://doi.org/10.15779/Z38HD7NT5B

© 2022 Sonia K. Katyal.

[†] Associate Dean for Faculty Development and Research and Haas Distinguished Professor, University of California at Berkeley School of Law. The author wishes to thank each of the wonderful authors who wrote papers for their insights and conversation, Pamela Samuelson, Paul Schwartz, Rebecca Wexler, and Angela Zhao, who offered excellent research assistance and support.

1. Joel R. Reidenberg, *Lex Informatica: The Formulation of Information Policy Rules Through Technology*, 76 Tex. L. Rev. 553, 554 (1998).
2. *Id.* at 554–55.
3. *Id.* at 554.
4. *See* Digital Millennium Copyright Act, H.R. 2281, 105th Cong. § 6 (1997).
5. *See generally* LAWRENCE LESSIG, CODE AND OTHER LAWS OF CYBERSPACE (1999).

Today, however, we confront a different phenomenon, one that requires us to draw upon the wisdom of Reidenberg's landmark work in considering the repercussions of the previous era. As much as Lex Informatica provided us with a descriptive lens to analyze the birth of the internet, we are now confronted with the aftereffects of decades of muted, if not absent, regulation. When technological social norms are allowed to develop outside of clear legal restraints, who wins? Who loses? In this new era, we face a new set of challenges—challenges that force us to confront a critical need for infrastructural reform that focuses on the interplay between public and private forms of regulation (and self-regulation), its costs, and its benefits.

This turn is undeniably significant, and it draws similarities to an earlier period in intellectual property history. In the 1980s and 1990s, a body of scholarship began to flourish that emphasized postmodern, feminist, critical race, and post-colonial approaches to intellectual property and information law.[6] These approaches, in part, focused their gaze towards how intellectual property entitlements facilitated the widening scope of inequality for some and also served as a source of empowerment for others.[7] Other scholars studied how even commons-like frameworks, seemingly open and free for all to use, actually benefited the powerful at the cost of disenfranchised groups. Madhavi Sunder and Anupam Chander, for example, authored a milestone article that pointed out how the concept of a public domain—and the romance attached to it—unwittingly contributes to a widening scope of inequality by offering a racialized libertarianism that celebrates appropriation, often at the cost of marginalized groups.[8] In addition, these works were also linked to the emerging field of Critical Information Studies (CIS), a field that focused on studying the abilities, rights, and limitations of the ways that users, consumers, or citizens alter or critique cultural texts, and the role of property rights and other forms of information control in limiting flows of information.[9]

---

6. Carys J. Craig, *Critical Copyright Law and the Politics of 'IP' in* RESEARCH HANDBOOK ON CRITICAL LEGAL THEORY 301, 304 (Emilios Christodoulidis, Ruth Dukes & Marco Goldoni eds., 2019).

7. *Id.*

8. *See* Boatema Boetang, *Symposium: Walking the Tradition-Modernity Tightrope: Gender Contradictions in Textile Production and Intellectual Property Law in Ghana*, 15 AM. U. J. GENDER SOC. POL'Y & L. 341, 345–46 (2007) (noting how cultural products of indigenous peoples are appropriated from the public domain and then repackaged as protected intellectual property, such as Ghanaian cloth designs); RUTH L. OKEDJI, CTR. INT'L GOVERNANCE INNOVATION, TRADITIONAL KNOWLEDGE AND THE PUBLIC DOMAIN 15–16 (2018) (arguing that the public domain benefits existing beneficiaries of the IP system and undermines creativity and innovation in local communities and indigenous peoples); K.J. Greene, *Intellectual Property at the Intersection of Race and Gender: Lady Sings the Blues*, 16 AM. U. J. GENDER, SOC. POL'Y & L. 365, 370–71 (noting how copyright law appropriated Black cultural production and made invisible the contributions of Black ragtime, blues, and jazz artists); Anjali Vats & Deidré A. Keller, *Critical Race IP*, 36 CARDOZO ARTS & ENT. L.J. 735 (2018) (outlining a scholarly movement that focuses on the racial disparities generated by the enforcement and ownership of intellectual property).

These earlier works, I would argue, have attained even greater salience in contemporary times, where the Black Lives Matter movement rightfully forces us to reckon with the need for a deeper interrogation—or perhaps integration—of the frameworks of social justice and intellectual property.[10] Instead of demonstrating the richness, complexity, and promise of yesterday's internet age, today's events show us what precisely can happen in an age of information libertarianism, underscoring the need for a new approach to information regulation. The articles in this Issue are taken from two separate symposiums—one on Lex Informatica and another on race and technology law. At present, a conversation between them could not be any more necessary. Taken together, these papers showcase what I refer to as the Lex Reformatica of today's digital age. This collection of papers demonstrates the need for scholars, lawyers, and legislators to return to Reidenberg's foundational work and to update its trajectory towards a new era that focuses on the design of a new approach to reform.

Below, I highlight five principle themes drawn from these collected articles that showcase the need for new generations of reform and regulation, i.e., what I refer to as today's 'Lex Reformatica.' The first concerns the need for infrastructural reform; the second involves a close attention to the negative impacts of underregulation; the third involves a focus on design and, relatedly, the concept of design justice; the fourth on reforming the interplay among public and private forms of regulation; and the final principle, which emphasizes the value of ex ante, instead of ex post forms of remediation.

---

9. Joel R. Reidenberg, *Lex Informatica: The Formulation of Information Policy Rules Through Technology*, 76 TEX. L. REV. 553, 554 (1998). See Siva Vaidhyanathan, Afterword: Critical Information Studies, A Bibliographic Manifesto, 20 Cultural Stud. 292, 293 (2006).

10. For more commentary on the intersection between intellectual property and social justice, see Lateef Mtima, *IP Social Justice Theory: Access, Inclusion and Empowerment*, 55 GONZ. L. REV. 401 (2020); HANDBOOK OF INTELLECTUAL PROPERTY AND SOCIAL JUSTICE (Stephen Jamar & Lateef Mtima eds., 2021); Peter Menell, *Property, Intellectual Property and Social Justice: Mapping the Next Frontier*, 5 BRIGHAM-KANNER PROP. RTS. CONF. J. 147 (2016); and Anupam Chander & Madhavi Sunder, *Is Nozick Kicking Rawls's Ass? Intellectual Property and Social Justice*, 40 U.C. Davis L. Rev. 563 (2007), along with the collection of essays in the Symposium.

## II. THE NEED FOR INFRASTRUCTURAL REFORM

At the heart of Reidenberg's insight was an analogy between the newly networked environment and the instability that early merchants faced in navigating differing jurisdictions.[11] For him, the management of content, personal information, and the preservation of ownership were three core areas of disruption. To navigate the uncertainties surrounding each domain, Reidenberg drew on early principles of Lex Mercatoria to argue that parties can carve out their own set of customs and practices, independent of local rules but which assured "basic fairness in their relationship."[12] Applying these principles to information technology, Reidenberg contended that information rules, i.e., rules of design, could do the same thing.

As several of these papers describe, Lex Informatica inspired a host of works that studied the norms, customs, and practices that characterized the growth of the internet, a world that emerged, initially, largely free from direct, legal regulation (and one that was considered superior for precisely this reason). Scholars were deeply skeptical of the need for or the benefits of regulation. Often, any discussion of regulation was immediately equated with overregulation. Consider one representative observation from this period from Lawrence Lessig: "Overregulation," he wrote, "stifles creativity. It smothers innovation. It gives dinosaurs a veto over the future. It wastes the extraordinary opportunity for a democratic creativity that digital technology enables."[13]

This view, widely shared by internet law academics at the time, aptly characterized the initial generation of internet-related scholarship, barely disguising a distrust of regulation and state overreach. In Lessig's single quote, we see, essentially, the union of three presumptive ideals in digital technology—first, the idea that regulation would stifle innovation; second, implicitly, that digital technology ruled by norms, rather than law, was preferable; and third, that digital technology enabled a democratization of creativity.

In a sense, *Lex Informatica* typified, and inspired this view. Yet if the initial growth of the Web's infrastructure might be characterized by the absence of direct regulation, law still played an indirectly powerful role in underscoring the trajectory of the Web's initial growth. And this is where our opening essay, from Rashida Richardson, becomes most salient. As Richardson has astutely observed, law does not exist in a vacuum—its background influence has contributed to a myriad of issues surrounding digital inequality.[14] These inequalities are intimately linked, both to the presence and absence of legal regulation. Consider, as she points out, the ways in which the history of de jure and de facto segregation in the United States has fed into the assembly and collection of training data, feeding algorithmic systems that generate AI-driven products that perpetuate racial inequality.[15]

---

11. Reidenberg, *supra* note 1, at 55–54.
12. *Id.* at 553.
13. LAWRENCE LESSIG, FREE CULTURE: HOW BIG MEDIA USES TECHNOLOGY AND THE LAW TO LOCK DOWN CULTURE AND CONTROL CREATIVITY 151 (2004).

Taking Richardson's work one step further and viewing it in another light, one can see how her powerful observations about the legacy of racial segregation requires us to think on an infrastructural level about the need for a critical analysis of data-driven technologies and the products that they create. As she astutely points out, segregation can reassert itself in a myriad of local, contemporary formations involving structural inequality.[16] This requires a closer examination of specific historical and contemporary laws, customs and social practices (norms) to see how bias can reassert itself.[17] And this interrogation, Richardson suggests, must take place at a local level.[18] Here, Richardson argues that a focus on technological injustice is incomplete without studying the historical practices that contribute to systemic inequality:

> These fields [focusing on legal procedures to create greater transparency or oversight] and the interventions they produce generally have two issues. First, they fail to reckon with the disadvantages and harms that preceded and are often compounded by data driven interventions. Second, they fail to decenter technology as the primary lens of analysis or modality of prevention and redress.[19]

The legal solutions to bias that are often promulgated, she points out, are mostly procedural in nature, she points out, and thus risk entrenching structural inequality.[20]

Threaded throughout Richardson's powerful article is the idea that in order to address algorithmic bias, we need to study the historical infrastructure that promotes structural inequalities. Only by decentering technology as the main lens of analysis and looking to the root cause of injustice, Richardson notes, can we begin to address algorithmic bias and other data-driven technologies that engender inequality. Here, too, Richardson offers us an intervention that focuses on infrastructural change: transformative justice. Her conclusion closes with a call towards employing a transformative justice framework, which she argues uses a systems-oriented approach that examines collective societal responsibility in creating systemic harms and centers people who are often "excluded from but pivotal to" the issues that data driven technologies raise.[21] Transformative justice, she argues, is necessary to create meaningful interventions to the problems of data-driven technology and advance society beyond the status quo.[22]

---

14. Rashida Richardson, *Racial Segregation and the Data-Driven Society: How Our Failure to Reckon with Root Causes Perpetuates Separate and Unequal Realities*, 36 BERKELEY TECH. L.J. 1051, 1053 (2021).
15. *Id.* at 104–6.
16. *Id.* at 106.
17. *Id.*
18. *Id.* at 108 (calling for clarity on the nature of the problem locally).
19. *Id.* at 135–36.
20. *Id.* at 136.

Richardson's call to excavate the infrastructure of historical inequality and transformative justice is nicely mirrored by Sandoval's excellent paper on ISP throttling, which also evokes a similar concern for infrastructural justice, or (as she calls it) "technology justice."[23] Whereas Richardson focuses on the historical framing of segregation and its contributions to algorithmic bias, Sandoval offers us a contemporary illustration of how structural and historical inequality can fuel a deprivation of access to critically important information. As she argues, "[i]nfrastructure regulation creates the future's physical and social architecture," pointing out that slowing down of access to the internet leaves users unable to access news sources, telemedicine, or to use videoconferencing necessary for work or school.[24]

While Richardson calls for a framing that focuses on transformative justice, Sandoval's elucidation of technology justice offers a similar, complementary reframing. She draws from the historical roots of the digital divide, showing us its contemporary aftereffects in the problem of ISP throttling. But even more presciently, she offers us a solution that focuses both on reforming the notion of transparency and the notion of internet access simultaneously:

> Inadequate disclosure in small faded print that does not make the consequences of ISP throttling clear is inconsistent with internet openness and may violate FCC transparency and FTC deceptive

---

21. *Id.* at 139–40.
22. *Id.* at 140.
23. *See generally* Catherine J.K. Sandoval, *Technology Law As A Vehicle For Technology Justice: Stop ISP Throttling To Promote Digital Equity*, 36 BERKELEY TECH. L.J. 963 (2021) (referring to "technology justice" in the title and various parts of the article).
24. *Id.* at 983.

> conduct laws and regulations. The prevalence of inadequate disclosures across carriers underscores the need for FCC and FTC regulatory action to protect consumers, internet openness, public safety, and the public interest.[25]

Sandoval, here, focuses on the notion of transparency but gives it a robust and active linkage to the idea of encouraging greater (and more meaningful) access to the internet. She recommends, first, exploring whether consumers are properly informed about ISP practices to enable consumer choice; and second, that the government collect more data on the existence of ISP throttling—and who it affects and how.[26] Finally, Sandoval argues, "[c]onsistent with corporate pledges to promote equity and inclusion, ending ISP practices that close the digital schoolhouse, healthcare, and economic opportunity door by throttling users back to the 90s would enable equity, inclusion, public health, and public safety."[27]

## III. THE IMPACT OF UNDERREGULATION

Sandoval's observation above brings us to the second principle of Lex Reformatica, drawn from a collection of these essays: reformers must recognize the negative impact of technology underregulation on individual civil rights, like privacy, due process, and equality. While many can remain anxious that needed reforms might imperil the freedom that originally defined the frontiers of cyberspace, it bears mentioning, as Gautam Hans has pointed out, that "the internet is already not functioning in so many obvious ways," pointing out that the current regime of decentralized, uneven regulation has produced troubling consequences.[28]

While Lex Informatica was written right after the dawn of the internet, we might view the papers by Tiffany Li, Gautam Hans, and Catherine Sandoval as a collection of studies that astutely demonstrate the negative impacts of a primary commitment to marketplace control. In a variety of circumstances, our commentators have shown us that the rise of technological norms, when unencumbered by close regulation of the marketplace, can flourish at the cost of equality, privacy, and due process.

Consider, for example, privacy law as an example of this trajectory. Reidenberg was known first and foremost as a privacy scholar. Had Reidenberg been alive today to see the degree of federal inaction in protecting privacy as a civil right, he would be fairly disappointed, particularly in light of his substantial European work affirming the link between privacy protection and regulation. As Tiffany Li has argued in her contribution to this Issue, privacy is a civil right and yet carries a kind of internal unevenness: as a body of principles, privacy law has failed to account for its *own* inequality in the sense that different people enjoy different levels of protection, both in type and intensity.[29] The more that we move into online spaces, Li points out, the more we are connected and the more opportunities there are for our civil rights to be violated in non-traditional ways.[30]

---

25. *Id.* at 132.
26. *Id.* at 133.
27. *Id.* at 134.
28. G.S. Hans, *Revisiting* Roommates.com, 36 BERKELEY TECH. L.J. 1227, 1250 (2021).

This is true, not just regarding privacy, but also regarding anti-discrimination as well. Take, for example, the case of *National Fair Housing Alliance v. Facebook*, discussed by Li.[31] In that case, Facebook permitted advertisers to selectively exclude racial segments of the population from viewing housing ads.[32] As Li points out, Facebook was able to discriminate against certain groups by relying on targeted advertising. Yet since targeted advertising relies intrinsically on the practice of data collection, Facebook's action was also a type of downstream harm affecting privacy because it only arises by virtue of the invasive data policies that enable racial categorization in the first place.[33]

Li's deft weaving of privacy and antidiscrimination concerns highlight a crucial distinction between privacy protections that function as civil liberties versus privacy protections that function as civil rights.[34] As she argues, "unequal access to privacy is a civil rights problem."[35] If we only envision privacy as a civil liberty, we miss the interplay between privacy and equality, doing a disservice to both realms.[36] In order to protect both, Li argues that we must reconceive of privacy as integral to both due process and equal protection:

> Laws and practices that promote surveillance, mandate the use of biased algorithmic assessment, and allow for gendered harms related to cyberstalking, should also be considered unconstitutional based on due process and equal protection. Individuals should be able to claim a constitutional right to privacy under the Equal Protection Clause, recognizing that privacy has never been awarded equally to all people across society.[37]

---

29. Tiffany C. Li, *Privacy As/And Civil Rights*, 36 BERKELEY TECH. L.J. 1265, 1269 (2021).
30. *Id.* at 114.
31. *Id.*
32. *Id.* at 115 (citing Complaint, Nat'l Fair Hous. All. v. Facebook, Inc., No. 1:18-cv-02689 (S.D.N.Y. Feb. 6, 2019)).
33. *Id.* at 115.
34. *Id.* at 108.
35. *Id.* at 109.
36. *Id.* at 110.

Li closes with a call for a federal privacy law that would function to help fill in the gaps that sectoral approaches have left behind, arguing that it would better situate privacy as a civil right, and put us on better footing with other privacy-forward nations, like the EU.[38]

A related kind of interplay, explored by Gautam Hans in his essay, *Revisiting* Roommates.com, details the relationship between the marketplace, speech protections, and inequality. Here, too, we see the human costs of a failure to integrate antidiscrimination protections into our efforts to regulate the internet. As has been widely discussed in technology law literature, Section 230 immunizes websites from liability for publishing unlawful speech that is made by third parties on their owned platform.[39] The reasoning behind this safe harbor is relatively straightforward: "[a]t the scale at which the platforms hope to operate," Hans explains, "the potential for liability would be immense, as would the costs of prescreening content."[40]

Yet in *Roomates.com*, two organizations filed suit against Roomates.com, contending that the company violated housing discrimination laws by mandating that end users fill out questionnaires that required disclosure of a user's age, gender, sexual orientation, and familial status, all of which are identity categories protected by federal fair housing laws.[41] In that case, the Ninth Circuit found that since Roomates.com facilitated connections between third parties (mandating that users answer questions that could result in discrimination), it could not be completely absolved from liability.

Yet Hans's intervention, however, imaginatively forces us to explore how *Roomates.com*, as an entity, would have fared in the absence of the protective sphere of Section 230. This move, in turn, asks us to situate *Roomates.com* alongside the history of housing discrimination and tenants' rights that characterizes the intersection of civil rights and housing discrimination.[42] Just as Li's paper asks us to imagine a broader framing of privacy law (and its attendant inequalities), Hans asks us to broaden our framing of equality principles in order to reimagine antidiscrimination protections among platforms. As Hans points out, the import of Section 230 essentially enables platforms to escape liability merely because of their online status, as opposed to newspapers, for example, which would have been required to comply with the Fair Housing Act.[43] But it's equally revealing, Hans points out, that civil rights concerns were excluded from its list of exceptions to immunity, an exclusion which has the less desirable effect of potentially creating an exception "large enough to potentially swallow" antidiscrimination protections entirely.[44] Hans concludes by calling for Section 230 reform that expands the role it could play in the civil rights and racial justice movements; supporting, rather than overruling, the goals of the Fair Housing Act.[45]

---

37. *Id.* at 124.
38. *Id.* at 125.
39. *See* Hans, *supra* note 28, at 103.
40. *Id.*
41. *Id.* at 107.
42. *Id.* at 105.

Again, both papers highlight the aftereffects of decades of federal underregulation, illustrating the negative externalities that affect individuals and their civil rights—regarding both privacy and equality.

## IV. THE DESIGN OF REFORM

A third principle in the Lex Reformatica landscape involves the concept of design as a stand-in for direct regulation. The notion of design-oriented solutions is a thread that weaves through many of the papers but is most explicitly explored by Hans (among others), who invokes design-oriented solutions as one way to remedy the inequalities that flow from immunity under Section 230. One solution to the quandary he explores involves a powerful refiguring of design choices—Section 230 could be reformed to create a potential opening for liability for design decisions (i.e., the choices a company makes in setting up its system)—such as its design of its drop down menus, pre-screening its content for liability issues, and the like.[46] In this way, a company would be responsible, ex ante, for ensuring that its design choices do not invite discrimination by its users.[47]

Hans' invocation of design reform, in many ways, indirectly echoes many of the same insights that have been associated with the emergent design justice movement, a concept associated with Sasha Costanza-Chock.[48] A general definition of the movement is "a field of theory and practice that is concerned with how the design of objects and systems influences the distribution of risks, harms, and benefits among various groups of people."[49] Here, special attention is paid to whether design reproduces (or is reproduced by) matrices of domination.[50] The concept of design justice is also oriented normatively, in that it works to build solutions that ensure fair and meaningful participation in design decisions and to recognize community based design and practice.[51] Overall the concept of design justice seeks to provide a "more equitable distribution of design's benefits and burdens," along with meaningful participation in design choices, and recognizes the value of community based practices.[52]

---

43. *Id.* at 113–14.
44. *Id.* at 116.
45. *Id.* at 122–25.
46. *Id.* at 122.
47. *Id.* at 123.
48. *See* Sasha Costanza-Chock, *Design Justice, A.I., and Escape From the Matrix of Domination*, J. DESIGN & SCI. (July 16, 2018), https://jods.mitpress.mit.edu/pub/costanza-chock.

Notably, Hans' treatment of design reform (and the association it indirectly draws with the notion of design justice) highlights yet another core aspect of Lex Informatica: the idea that "law is not the only source of rules or rulemaking."[53] As Margot Kaminiski explains, drawing from Reidenberg: "[t]echnological architecture is its own distinct regulatory force. This insight has serious implications for the law. It means…. That technology isn't understood to be value-neutral, authoritative, or inevitable. It reflects choices. It's political."[54]

As she argues, Reidenberg was "the first to say that architecture mattered."[55] But the interaction, she points out, between law and technology does not have one singular formation; it can take on a variety of different formations. Here, law can structure the development of certain technologies, be tied closely to policy goals and values, or make salient particular aspects of doctrine.[56] Law, in this sense, operates to construct technology into its own systems of meaning and value.[57] "The internet isn't a no-lawyer's land; it's populated by people who need stability, norms, rules, and consequences. People *need* the law and *grow* the law; it isn't imposed upon them."[58]

---

49. Sasha Costanza-Chock, *Design Justice: Towards an Intersectional Feminist Framework for Design Theory and Practice*, in 2 DESIGN AS A CATALYST FOR CHANGE–DRS INT'L CONFERENCE 2018 529, 529 (Cristiano Storni, Keelin Leahy, Muireann McMahon, Peter Lloyd & Erik Bohemia eds., 2018); *see also* Sasha Costanza-Chock, DESIGN JUSTICE: COMMUNITY-LED PRACTICES TO BUILD THE WORLDS WE NEED (2020).
50. *Id.* at 533.
51. The concept of design justice grew out of a summit of designers, artists, technologists, and community organizers in a 2015 meeting of the Allied Media Conference. *Id.* at 529.
52. *Id.* at 533.
53. Margot E. Kaminski, *Technological "Disruption" of the Law's Imagined Scene: Some Lessons from* Lex Informatica, 36 BERKELEY TECH. L.J. 883, 886–887 (2021).
54. *Id.*
55. *Id.*
56. *Id.* at 110–12.
57. *Id.* at 112.

While Reidenberg recognized that law's reach would be inevitable over the internet's inherent messiness, Kaminski subtly exhorts us to return to the messiness of technology in order to excavate how technology changes the "imagined regulatory scene," or the law and policy conversations that are imagined to take place.[59] For Kaminiski, the rise of technology and its architecture can disrupt these imagined scenes, forcing us to examine and reconstruct the nature and justification of regulation itself—as she writes, "not just the 'how' and 'what' of law, but also the 'why.'"[60] As she points out, the very meaning of technology takes on a certain particularity in tandem with "what one thinks the law is or should be"; in other words, the law constructs technology, just as it constructs other categories of meaning.[61] Here, we should interrogate the balancing of these values and categories, asking constantly whether or not we have struck the correct equilibrium and exposing the particular values that we are acting upon.[62]

## V. CRITICAL EXAMINATION OF THE INTERPLAY OF PUBLIC AND PRIVATE REGULATION

Kaminiski's description of the law's "imagined scene" brings us nicely to the fourth aspect of Lex Reformatica: the importance of critically examining the interplay between public and private forms of regulation. Whereas Reidenberg's work was characterized by a clear dividing line between public and private forms of architectural regulation, our more modern era has revealed that there are substantial slippages and boundaries between these forms. Private forms of regulation permeate the internet, as content moderation and other strategies of compliance have flourished in the last thirty years. At the same time, just as Reidenberg implicitly predicted, these private forms of regulation have emerged, just as public forms of regulation have largely faded into the background of technology, until just recently.

This is not entirely by accident. As Julie Cohen describes in this Issue—in comparing Lessig's *Code* to Lex Informatica, one is struck by the divergence of their approaches.[63] As she explains, because of Reidenberg's exposure to North American and European ways of approaching law and regulation, he focused his gaze on the mechanics of integrating regulatory authority into the development of technologies.[64] In contrast, *Code*, being a product of Lessig's training at the University of Chicago, focused much more on the role of social norms and markets in influencing the design of technologies.[65] She writes:

> And yet "Lex Informatica," but not *Code*, surfaced the complex *interplay* between regulatory forces. "Lex Informatica" framed new digital formations as situated opportunities for interventions by policymakers and other interested actors—an approach broadly compatible with decades of accumulated, interdisciplinary learning on emergent sociotechnical processes—whereas *Code* described an elemental regulatory struggle that unfolded as a contest over *terra nullius* and that resonated with the reigning neoliberal ethos of the era.[66]

---

58. *Id.* at 106.
59. *Id.* at 104.
60. *Id.* at 131.
61. *Id.* at 112.
62. *Id.* at 131.

As Cohen eloquently observes, both *Code* and Lex Informatica were ill-equipped to handle the unexpected developments that came with the evolution of the relationship between law and technology.[67] For example, how does one define "compliance" in an age of algorithmic processes?[68] What legal obligations do platforms face in addressing the activities of their end users?[69] How do we ensure public accountability over compliance operations, Cohen asks.[70] How do we ensure citizens' privacy and dignity in a world of data-driven surveillance?[71]

This brings us to a fourth element in the architecture of an era of Lex Reformatica: a critical focus on the interplay of public and private forms of regulation (and self-regulation). In today's contexts, we are grappling with the rise of private compliance systems, systems that are designed to both satisfy legal parameters at the same time that they can often enable circumvention of civil rights principles. In Cohen's masterful exploration of organizational transformation, she argues that "networked information technologies are not simply new modes of knowledge production to be governed, but also powerful catalysts for organizational restructuring that change the enterprise of governance . . . from the inside out."[72]

---

63. Julie E. Cohen, *From Lex Informatica to the Control Revolution*, 36 BERKELEY TECH. L.J. 1017, 1019 (2021).
64. *Id.*
65. *Id.*
66. *Id.*
67. *Id.*
68. *Id.*
69. *Id.*
70. *Id.* at 104.
71. *Id.*

In the context of platform speech, for example, Cohen brilliantly elucidates how a reliance on probabilistic and engagement metrics in online communities drives users toward extremist speech, essentially foreclosing the efficacy of content or speaker interventions to disrupt their mechanisms.[73] Among copyright conflicts and otherwise, speech and content platforms have emerged and evolved, elevating generative speech and content over more rigorous forms of gatekeeping.[74] And under these circumstances, Cohen argues, civil rights advocates are stuck, unsuccessfully trying to bargain with tech giants who wrap many of their activities within a thick shroud of secrecy, foreclosing mutually acceptable forms of compromise.[75] Not surprisingly, amidst this climate, a variety of kinds of privatized governance emerge, producing "vast new compliance industries dedicated to the pursuit, perfection, and legitimation of self-governance."[76] While Cohen's description might lead us to consider the failures and limitations of these efforts, she ends on a note that encapsulates Reidenberg's persistent optimism: center innovation in the law; consult private industry but remain skeptical, that is, avoid equating self-interested positioning with the importance of human flourishing; and finally, "remember that law is a means to an end."[77]

Whereas Cohen offers us a broad, abstract view of the relationship between private and public forms of regulation, Deirdre Mulligan and Ken Bamberger take a narrower, functional approach in their study of content moderation. Here, the authors bravely—and deeply—engage with the porous and shifting strands of public and private content moderation, arguing that it involves distributed forms of public and private oversight that are, in turn, mostly delegated to a diverse range of actors. Drawing on a range of case studies—the DMCA, the General Data Protection Regulation (GDPR), the governance of online material tied to child sexual abuse, Section 230, and the right to be forgotten—revealing how content moderation tools delegate and constrain decision-making by private actors through deploying a typology of subfunctions.[78] By looking to these actions, which they describe as *defining*, *identifying*, *locating*, and *moderating*, Mulligan and Bamberger then use these subfunctions to pose a deeper interrogation of these forms of private governance in light of the public values of accountability.[79]

---

72. *Id.* at 111.
73. *Id.* at 113–14.
74. *Id.* at 117–18.
75. *Id.* at 120–22.
76. *Id.* at 125.
77. *Id.* at 134.

Drawing in part on a previous collaboration with Helen Nissenbaum and the work of the New Governance school of thought, which focuses on the decentering of public forms of governance by private actors,[80] the authors argue that in order to truly understand the values promulgated by content moderation systems, we need to identify, separate, and examine the various subfunctions that operate in content moderation and the various hand-offs that take place between human and technical means.[81] In applying this insight to a broad array of mechanisms, all of which focus on content moderation, the authors beautifully lay out the various critiques associated with each form and the ethical and political questions that arise.[82] To ameliorate some of the disadvantages of these subfunctions, particularly regarding various degrees of definitional competency, the authors profitably argue in favor of a kind of transparency that takes into account the need to disclose the definitions and decisional criteria used in content moderation, and to foster the participation of other stakeholders.[83] Other considerations that they describe involve developing more competencies through focusing on the cultural contexts that surround content moderators and their fit with the content that is being regulated.[84]

As Mulligan and Bamberger demonstrate, a close eye to the intricacies of the functions—and subfunctions—of content moderation can yield important insights into the risks and benefits behind private forms of content moderation. But it is important to note, as Hanna Bloch-Wehba points out in her essay, the porosity of the relationship between public and private, perhaps unintentionally, creates a situation, the extent to which forces us to confront the effect of content moderation on law enforcement and vice versa. As she argues, "the purportedly private rules of content moderation are created and

---

78. Deirdre K. Mulligan & Kenneth A. Bamberger, *From Form To Function in Content Moderation*, 36 BERKELEY TECH. L.J. 1091, 1110 (2021).
79. *Id.* at 106.
80. *Id.* at 115.
81. *Id.* at 115–16 (citing Deirdre K. Mulligan & Helen Nissenbaum, *The Concept of Handoff As a Model for Ethical Analysis and Design*, in OXFORD HANDBOOK ETHICS & AI 234 (Markus D. Dubber, Frank Pasquale & Sunit Das eds., 2020).
82. *Id.* at 107.
83. *Id.* at 153.
84. *Id.* at 156.

operate within a political context in which law enforcement acts as a particularly powerful stakeholder."[85]

This produces a dialectic between law enforcement and social media content that constructs both trajectories in turn: content moderation, i.e., speech surveillance on platforms, shapes the scale and design of law enforcement and vice versa. "Just as law enforcement seeks expanded influence over platforms' private decision-making, the processes and technical affordances of content governance also affect and shape law enforcement investigations in more mundane contexts."[86] Here, while law enforcement aims to influence content moderation on platforms, the substantive, procedural, and technical rules that govern platforms shape law enforcement itself.[87] She studies the complex relationship between government pressure to moderate content and private platforms' responses, noting how government influence can shape private content moderation systems.[88] In turn, the private decision-making (including standard setting) activities of firms can also shape the behavior of law enforcement in seeking data and user information, even arguably making it more difficult to procure content involving illegal activity.[89]

In making these observations, Bloch-Wehba offers another crucial variable for consideration in the discussion of private and public forms of regulation: the self-interested nature of law enforcement. As she brilliantly elucidates, the technological modalities that govern online content can also feed the interests of law enforcement, creating "new types and sources of information relevant to new kinds of investigations."[90] In contexts as varied as terrorist content and sex work, private platforms essentially operate as proxy censors; even though platforms enjoy some modicum of immunity, they are still indirectly pushed by law enforcement to behave more aggressively in filtering unlawful content.[91] As she explains:

> In the context of terrorist imagery, platforms are required to report certain kinds of terroristic threats to European authorities and likewise required to preserve a broader range of information for future law enforcement use. The result is that the monitoring technology used to detect lawbreaking itself lies at the heart of investigations and prosecutions, yielding increasing entanglements between law enforcement and platform governance.[92]

---

85. Hannah Bloch-Wehba, *Content Moderation as Surveillance*, 36 BERKELEY TECH. L.J. 1297, 1300 (2021).
86. *Id.*
87. *Id.* at 106.
88. *Id.* at 110–11.
89. *Id.* at 105.
90. *Id.* at 106.
91. *Id.* at 117.

The creeping shadow of law enforcement over these privatized forms of governance, Bloch-Wehba points out, forces us to confront how the implicit, indirect collusion between the two dismantles the prevalent assumption that private platforms wholly engage in self-governance and are accountable to no one but their shareholders.[93] The "extensive alignment" of platforms and law enforcement lacks the kind of accountability that our legal system is premised upon; twenty-three years after Lex Informatica, Bloch-Wehba notes, "U.S. law has made little progress in ensuring that *lex informatica* is as democratically legitimate or accountable as its regulatory equivalents."[94] Hence the need for a more critical, and searching, inquiry into future possibilities for its governance.

## VI. EX ANTE REGULATION VS. EX POST REMEDIES

Much of the critical points made by our authors in this Issue have pointed their gaze toward the interplay between private and public, pointing out (as Bloch-Wehba has done) how various incentives and interests can often complement one another, often at the cost of accountability to the public. In order to address this issue, Reidenberg himself advocated "a shift in the focus of government action away from direct regulation and toward indirect influence" by regulating behavior and standards before even the consequences of technology surface.[95]

This important insight about the value of proactive intervention, rather than after-the-fact forms of correction, operates at the heart of many of the essays in this Issue, but it appears most directly in Ifeoma Ajunwa's powerful study of the increasing use of automated video interviewing and the issues that it raises for employment discrimination.[96] She points out that these technologies are often touted as anti-bias interventions but can paradoxically run the risk of not just replicating the bias they are meant to evade, but also amplify it.[97]

Ajunwa's insightful case study threads a number of themes that I've discussed regarding the collection of essays, but her work demonstrates yet an additional principle: the value of ex ante legal regulations as opposed to ex post remedies.[98] As Ajunwa has noted in her application of Reidenberg, one critical benefit of Lex Informatica is that it relies on ex ante measures of execution.[99] Instead of remedying harm that has already occurred, Lex Informatica enables automated monitoring and enforcement even before the violation has even taken place.[100] Here, she draws directly on Reidenberg to show how thoughtful, proactive interventions at the design stage can radically deter discrimination at a later stage.[101]

---

92. *Id.* at 133–34.
93. *Id.* at 141.
94. *Id.* at 141, 144.
95. Ifeoma Ajunwa, *Automated Video Interviewing as the New Phrenology*, 36 BERKELEY TECH. L.J. 1173, 1218 (2021) (citing Reidenberg, *supra* note 1, at 586).
96. *See generally id.*
97. *Id.* at 103.

Ajunwa's article, here, is a masterful example of the risks and benefits of the AI-driven era we inherit today. Consider the case study that she relates, so vividly: (1) a new technology is invented (here, automated video interviewing) that seems to increase efficiency but that actually perpetuates bias; and (2) existing legal principles make it difficult to establish a case of illegal discrimination under existing protections (Title VII, the ADA, and various informational privacy entitlements). In such situations, Ajunwa argues that a Lex Informatica framework might push a more proactive intervention at a much earlier stage: "Whereas traditional law would require candidates to know a violation of their rights occurred in order to seek protection—a serious problem given the opaque nature of algorithmic decision-making—technological solutions under a Lex Informatica framework provide some assurance that such violations will not occur in the first place."[102]

Again, by enlisting design principles at an early stage, Ajunwa invocates Lex Informatica, arguing that the Uniform Guidelines on Employee Selection Procedures should be employed in the design of such systems, including the collection, validation, and use of particular content.[103] She even argues that Lex Informatica might provide the basis for a property entitlement in the subject's informational privacy interests, again showing us how an ex ante design orientation might foreclose the risk of bias and discrimination at a later stage.[104]

Ajunwa's insightful case study offers us a lesson in employing Lex Informatica in the very fraught area of recruitment and employment. While she shows us how a regulatory framework could be employed at the front end to govern and confront the potential bias that surfaces from one form of a new technology, in our final paper in this Issue, Karni Chagal-Feferkorn and Niva Elkin-Koren focus on a similar pattern emerging from a much broader array of technologies that they collectively refer to as Lex AI.[105] While there has been much ink spilled describing the domain of AI, the authors treatment reframes our gaze towards addressing not the issue of solving the problems of AI but rather *how* AI governs human behavior.[106]

---

98. *Id.* at 104.
99. *Id.* at 146–47.
100. *Id.*
101. *Id.* at 104.
102. *Id.* at 146–47.
103. *Id.* at 147–48.
104. *Id.* at 151–53.

Refreshingly, they argue, Lex Informatica has given way to another system of private ordering through technology, a system that includes personalized recommendations and other forms of data-driven decision-making.[107] Yet here, the authors stop short of describing Lex AI as just another form of private ordering; instead, they argue that it comprises an enabler of collective action.[108] The paper, like those mentioned in the Section above, carefully revisits the public/private distinction, especially in law and economics literature, which is often deployed to justify the appropriate scope of regulatory intervention. Yet here, the authors perceptively ask how to approach governance in such circumstances, particularly given its infrastructure which might seem (at first glance) to aggregate the will of connected individuals, but in fact (the authors point out) functions as a robust, unaccountable, mechanism that actively constructs, rather than collects, the will of individuals.

Unlike many of the other authors in this Issue, who undertake a notably critical gaze towards AI and its aftereffects, Chagal-Feferkorn and Elkin-Koren choose to instead consider Lex AI as a kind of sui generis "unicorn" type of governance: one that invites closer scrutiny because it lacks some of the typical advantages of private ordering, while also typifying some of the problems of collective action.[109] The authors draw an insightful comparison between Lex Informatica and decision-making by AI, arguing that the latter could be viewed as a system of governance because of the way that it generates norms and affects the behavior of users but that it can also be deployed by private or public entities.[110] Here, Lex AI also enables a greater degree of personalization in a variety of different areas, including decision-making in both the private and public spheres.[111]

---

105. Karni A. Chagal-Feferkorn & Niva Elkin-Koren, *Lex AI: Revisiting Private ordering by Design*, 36 BERKELEY TECH. L.J. 915, 919 (2021).
106. *Id.* at 107.
107. *Id.*
108. *Id.* at 108.
109. *Id.* at 109.
110. *Id.* at 117–18.
111. *Id.* at 121–22.

Perhaps the biggest payoff from their thoughtful account lies in their unwillingness to analogize Lex AI to a complete form of either public or private ordering. As they argue, while Lex AI "can be easily mistaken for a private ordering form of governance, [Lex AI] is in fact closer in nature to public ordering." But at the same time, the analogy is incomplete; as they show, Lex AI also lacks some of the key characteristics at the basis of public ordering, as well. AI drives its predictions from a centralized process of decision-making, leading the authors to conclude that it may resemble a distinctive type of collective action mediated by algorithms, rather than by self-governance.[112] As a result, Lex AI cannot qualify as strictly top-down governance, since it is formed by data-driven predictions, which are dynamic, distributed, and far less predictable than traditional modes of "command-and-control governance."[113]

Critically, the authors are careful to note that Lex AI is permeated by a particular type of power asymmetry. At the same time that it offers efficient ways to manage and analyze data, Lex AI cannot always reflect a user's personal choice.[114] The authors note, "Lex AI does not provide a reliable signaling of people's preferences and choices. The way preferences are inferred and the recursive process by which Lex AI shapes norms and behaviors may result in predictions that fail to reflect individuals' true preferences and may generate inefficiencies."[115]

In the end, the authors argue that the advent of Lex AI is wholly different than traditional forms of governance—it is not exactly centralized, nor is it totally distributed.[116] And this insight—one might characterize it as a refusal to analogize—offers a host of possibilities for a new way to approach governance. The authors close by suggesting that public policy "treat Lex AI as an ecosystem [shaped by] various types of (sometimes unidentified) entities," each of which carries the capacity to shape the adaptive learning of the centralized system, thereby affecting the decisions and predictions that it produces.[117] By including a broader array of stakeholders and by questioning the validity of data-related and design choices, Lex AI can better perfect the regulatory and legal measures that it may generate.[118]

---

112. *Id.*
113. *Id.*
114. *Id.*
115. *Id.*
116. *Id.*
117. *Id.* at 145–46.
118. *Id.*

Of course, the central question that threads through their article, along with each of those I've discussed in this Foreword, involves the question of governance. Because of its data-driven, dynamic nature, the authors point out that ex ante scrutiny of its systems may be simply impossible to perform.[119] Because its adaptive functions are driven by data, they are inherently dynamic; the norms that embed the system are opaque and nondiscursive, short-circuiting the opportunity for an interrogation of its values. Even more than its opacity, or in part because of it, Lex AI fails to facilitate a discursive engagement into how it functionally corresponds to social norms and values and thus fails to enable a deeper public deliberation of its utility.[120]

In the end, as they gently warn, our present inability to grapple with a system of governance of Lex AI may result in at least a partial undoing of—or at least a challenge to—democracy. But as they point out, the future does not necessarily have to be this way. Indeed, as they describe at the end of the essay, perhaps AI may result in a redoing (as opposed to an undoing) of democracy, if we can reconfigure our thinking:

> Treating Lex AI as a data ecosystem rather than a rule-based design would assist in focusing attention on the interaction between the different actors when considering legal tools to mitigate potential harms the system generates.[121]

For these authors, the mitigation of AI-related harms requires us to train our lens towards a wider gaze; an inquiry that considers more than code alone, and also reckons with the potential limitations inherent in the sources of data, as well as its recursive and dynamic effects on decision-making and governance.[122] As they warn, under this framing of an ecosystem, "design and deployment decisions pertaining to the system might be affected by different stakeholders" but should not be interpreted to function as a shield from liability for those choices.[123] Indeed, it is the opening of these possibilities—design choices, deployment choices, the interaction between stakeholders and the potential for liability for harm—that carries the greatest ex ante potential for improving regulatory and legal measures in the future.[124]

---

119. *Id.* at 145.
120. *Id.*
121. *Id.*
122. *Id.* at 147.
123. *Id.* at 148.
124. *Id.*

## VII.   CONCLUSION

These articles, as insightful and varied as they are, collectively represent a necessary conversation between the wisdom of prior generations of technology scholars like Joel Reidenberg, who focused on the possibilities of fairness through customs, norms, and the design of technology; and a newer generation of technology scholars who rightfully draw our attention to the aftereffects of an absence of regulation on vulnerable groups. Collectively, these papers represent an unfolding conversation about technology, norms, design, and regulation—a world of Lex Reformatica that Reidenberg himself would have deemed full of possibilities for meaningful integration in the future.